\documentclass[conference]{IEEEtran}
\IEEEoverridecommandlockouts

\usepackage{cite}
\usepackage{amsmath,amssymb,amsfonts}
\usepackage{algorithmic}
\usepackage{graphicx}
\usepackage{textcomp}
\usepackage{xcolor}

\usepackage[hyphens]{url}
\usepackage{enumitem}
\usepackage{booktabs}
\usepackage{tabularx}
\usepackage{booktabs}
\usepackage{multirow}
\usepackage{amsmath}

\newcommand{\rev}[1]{\textcolor{black}{#1}}

\begin{document}

\title{\textsc{PipeWeave}: Synergizing Analytical and Learning Models for Unified GPU Performance Prediction}

\author{
    \IEEEauthorblockN{
        Kaixuan Zhang\textsuperscript{1,3},
        Yunfan Cui\textsuperscript{1},
        Shuhao Zhang\textsuperscript{1},
        Chutong Ding\textsuperscript{1},
        Shiyou Qian\textsuperscript{1,*},
        Luping Wang\textsuperscript{2,*},
        Jian Cao\textsuperscript{1},\\ 
        Guangtao Xue\textsuperscript{1},
        Cheng Huang\textsuperscript{2},
        Guodong Yang\textsuperscript{2}, and
        Liping Zhang\textsuperscript{2}
    }
    
    \vspace{1.2mm}
    \IEEEauthorblockA{
        \textsuperscript{1}Shanghai Jiao Tong University, China \qquad
        \textsuperscript{2}Alibaba Group, China \qquad
        \textsuperscript{*}Corresponding authors \\
        \textsuperscript{3}Work done during an internship at Alibaba Group \\
        \{zks1anx, cuiyunfan, zhang-shuhao, qshiyou\}@sjtu.edu.cn
    }
    \thanks{Preprint version. Accepted to ISCA 2026.}
}


\maketitle

\thispagestyle{plain}
\pagestyle{plain}

\begin{abstract}
The rapid expansion of Transformer-based large language models has dramatically increased the need for high-performance GPUs. As a result, there is growing demand for fast, accurate, and widely generalizable GPU performance models to support next-generation hardware selection and system-level exploration. However, current data-driven methods are limited, exhibiting poor  generalization across hardware and inadequate modeling of complex production-level kernels common in modern inference stacks. To address these issues, we present \textsc{PipeWeave}, a unified GPU modeling framework. This approach first employs an analytical model to quantify a given kernel's demands on the GPU's heterogeneous instruction pipelines. These analytical features are then fed into a machine learning (ML) model to capture complex cross-pipeline interactions and resource dependencies, enabling high-fidelity performance prediction. Our evaluation across 11 GPU types from four generations of major architectures on two widely-used serving systems demonstrates that \textsc{PipeWeave} delivers high fidelity and strong generalizability. It achieves accurate predictions, with only 6.1\% average error at the kernel level and 8.5\% for end-to-end inference—reducing the error of state-of-the-art methods by 6.7$\times$ and 4.4$\times$, respectively. We also demonstrate \textsc{PipeWeave}'s value ``beyond simulation'' by utilizing its performance ceiling to diagnose implementation shortcomings and guide the optimization of a production fused MoE Triton kernel, achieving up to 1.7$\times$ speedup. Code is available \url{https://github.com/zksainx/pipeweave}.

\end{abstract}

\section{Introduction}
\label{sec:intro}

The advent of Transformer-based~\cite{transformer} large language models (LLMs) has fundamentally reshaped the landscape of artificial intelligence. Today, a vast array of LLMs—ranging from proprietary flagships like Gemini~\cite{gemini25} to open-source series like Llama~\cite{llama} and Qwen~\cite{qwen}—are deployed to power diverse services, from coding assistants to document summarization. Efficiently serving these varied workloads across heterogeneous hardware platforms requires large-scale clusters of high-performance nodes.

The surging demand for performance is met by a relentless pace of hardware innovation from major vendors like NVIDIA and AMD, who frequently release new GPU architectures with substantial performance and feature updates. For example, since the introduction of its Ampere architecture, NVIDIA has released \textbf{four} distinct architectures and \textbf{dozens} of non-consumer model variations targeting different market segments~\cite{cuda_gpus}. This rapid co-evolution of models and hardware poses a critical challenge for system designers and architects. The sheer volume of hardware configurations makes exhaustive testing impractical, while the inability to acquire every possible configuration—or even access unreleased, next-generation hardware—further compounds this issue. Therefore, to facilitate large-scale system-design, hardware selection, and the development of next-generation systems, the need for fast, accurate, and generalizable GPU performance models has never been more pressing.

Historically, GPU performance modeling has followed three main paradigms, each exhibiting a different trade-off among fidelity, speed, and generality. First, cycle-accurate simulators~\cite{gpgpusim,accelsim,mgpusim} offer the highest fidelity by emulating microarchitectural behavior in detail, but their simulation speed is computationally expensive and their lack of portability makes generalizing to new or undocumented hardware challenging. Second, analytical models~\cite{AMALI,GCOM,GPUMech} provide much faster estimates by relying on performance formulas such as interval analysis, yet their accuracy is often constrained and their dependence on hardware-specific microbenchmarking restricts generality to unseen architectures. Third, data-driven approaches~\cite{habitat,neusight} achieve high speed by learning tile-level latency from measurements, but their predictive accuracy can be variable and their high-level modeling assumptions—treating tiles as atomic, assuming uniform SM behavior, and not fully capturing fused-kernel coupling (e.g., FlashAttention~\cite{flashattention})—can impact generalization across workloads and hardware generations.


To bridge this gap, we present \textsc{PipeWeave}, a framework that achieves high fidelity, fast speed, and broad generalizability in GPU performance modeling through a combined analytical–ML design that \textit{weaves} pipeline-level analysis into accurate predictions. The framework first decomposes a given kernel into a set of fundamental \textit{tasks}, each representing a schedulable unit of work for a Streaming Multiprocessor (SM). It then simulates how these tasks are mapped onto SMs according to the kernel’s execution paradigm, producing a realistic task distribution. Based on this distribution, \textsc{PipeWeave} analytically derives each task’s \textit{pipeline demand} and associated \textit{theoretical cycles} for the SM’s heterogeneous instruction pipelines, and aggregates them into a compact multi-level feature set. Finally, a lightweight MLP consumes these features to predict the kernel's execution duration.

We conducted extensive evaluations to validate our framework. Our experimental testbed spans 4 hardware generations, encompassing 11 distinct GPU types (6 for training, 5 for unseen testing), and 5 categories of critical kernels (e.g., GEMM, Attention) in FP8, BF16/FP16, and FP32 precisions, commonly invoked by frameworks like vLLM\cite{vllm} and SGLang\cite{sglang}. At the kernel level, \textsc{PipeWeave} achieves a low average MAPE of 6.1\% on seen GPUs and 11.4\% on unseen GPUs, drastically outperforming the state-of-the-art (SOTA) baseline, Neusight~\cite{neusight}, representing an error reduction of 6.7$\times$ and 3.8$\times$, respectively. We further validate our model on complex inference workloads, using three large models (Qwen2.5-14B, Qwen3-32B, Llama3.1-70B) with various Tensor and Pipeline Parallel (TP/PP) configurations. In these E2E scenarios, \textsc{PipeWeave} maintains high fidelity, achieving an average error of 8.5\% on seen GPUs and 10.7\% on unseen GPUs—reducing the prediction error of Neusight by 4.4$\times$ and 3.1$\times$. Finally, we demonstrate \textsc{PipeWeave}'s value ``beyond simulation.'' By utilizing the model to establish a potential performance ceiling, we identify hardware-specific implementation inefficiencies in a production Fused MoE Triton kernel and guide targeted optimizations, achieving up to a 1.7$\times$ speedup.

In summary, this paper makes the following contributions.
\begin{itemize}
    \item \textbf{Unified Modeling Framework:} We propose PipeWeave, a unified framework synergizing analytical modeling with machine learning to accurately capture complex pipeline interactions for high-fidelity prediction.

    \item \textbf{Superior Generalization:} Validated across 11 GPUs spanning four generations, \textsc{PipeWeave} achieves SOTA accuracy on unseen architectures, reducing prediction error by up to 6.7$\times$ over prior methods.

    \item \textbf{Optimization Guidance:} We demonstrate utility ``beyond simulation'' by establishing performance ceilings to diagnose implementation inefficiencies and guide targeted optimization for production kernels.
\end{itemize}

\section{Background}

\subsection{Large Language Models (LLMs)}

Modern LLMs are predominantly based on the Transformer architecture~\cite{transformer}. A typical Transformer block includes two key components: a multi-head attention mechanism and a position-wise feed-forward network (FFN). 
LLM inference is commonly divided into two phases: \textit{prefill} and \textit{decode}~\cite{vllm,sglang,sarathi}. During prefill, the input prompt is processed in parallel, with key/value pairs computed and stored in the KV cache for every token. The decode phase then generates output tokens in an autoregressive way.

\begin{table}[tb]
    \centering
    \caption{\rev{Runtime breakdown of Qwen2.5-32B inference on a 4$\times$A100 cluster with Tensor Parallelism (TP=4).}}
    \label{tab:kernel_breakdown}
    \vspace{-2mm}
    \resizebox{\linewidth}{!}{
    \begin{tabular}{lcccccc}
        \toprule
        \textbf{Phase} & \textbf{GEMM} & \textbf{Attention} & 
        \textbf{RMSNorm} & \textbf{SiLU\&Mul} & \textbf{All-Reduce} & \textbf{Other} \\
        \midrule
        Prefill        & 72.70\% & 8.22\%  & 3.85\% & 2.26\% & 12.10\% & 0.87\% \\ 
        Decode         & 65.05\% & 17.78\% & 3.19\% & 1.50\% & 5.76\%  & 6.72\% \\ 
        \bottomrule
    \end{tabular}
    }
    \vspace{-4mm}
\end{table}


\rev{Table~\ref{tab:kernel_breakdown} shows the kernel (a GPU-executable function) runtime distribution for Qwen2.5-32B inference on a 4$\times$A100 cluster with Tensor Parallelism (TP=4) (batch size 8, sequence length 8192). These categories correspond to core computational building components and communication primitives of the Transformer architecture used in today's mainstream distributed LLMs: GEMM kernels~\cite{sarathi} dominate the workload, stemming from linear projections in both attention and feed-forward layers; Attention kernels compute the relationships between tokens; RMSNorm kernels~\cite{rmsnorm} stabilize activations prior to attention and feed-forward computations; operations like SiLU\&Mul implement activation functions and element-wise calculations, which are central to the SwiGLU FFN used in many LLMs~\cite{shazeer2020glu}; and All-Reduce kernels handle the essential collective communications across GPUs. Together, these major kernels account for the vast majority of the total runtime. As their dominance persists across models, software stacks, and hardware generations~\cite{sarathi,flashattention, flexgen}, our analysis focuses on accurately modeling these kernels.}


Furthermore, recent advances in LLM optimization have led to the adoption of specialized kernels. The use of lower-precision data types, notably FP8 quantization for W8A8 (8-bit weights and activations) inference~\cite{kuzmin2022fp8_quantization,shen2023efficient_fp8}, has popularized Scaled Matrix Multiplication (Scaled MM) kernels~\cite{cutlass_scaledmm, transformerengine_docs, deepgemm_repo}. Concurrently, Mixture-of-Experts (MoE) architectures~\cite{shazeer2017moe, fedus2022switch,dai2024deepseekmoe} leverage Fused MoE kernels. These kernels efficiently execute batched GEMM operations across expert sub-networks once token routing is finished.

\subsection{From Kernels to GPU Architectures}
\label{sec:GPU-arc}

\begin{figure}[tb]
    \includegraphics[width=0.95\columnwidth]{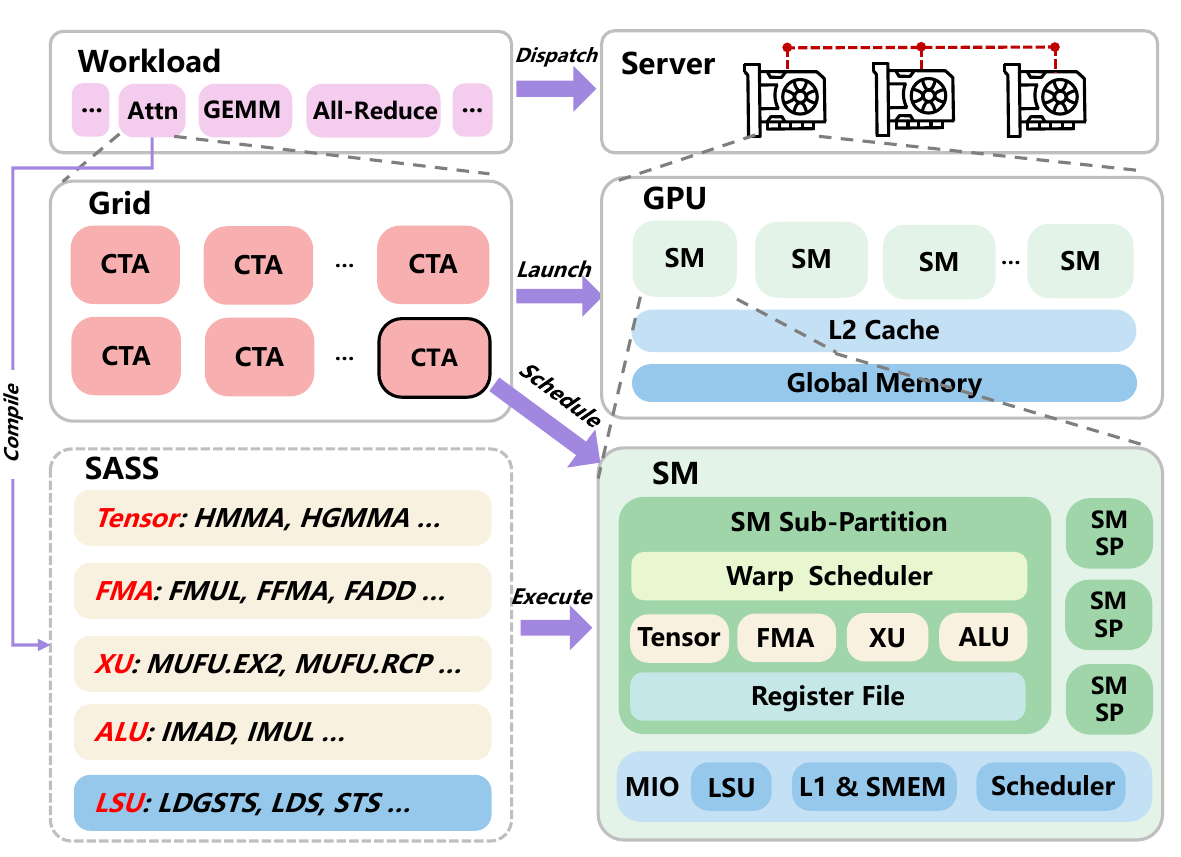}
    \vspace{-4mm}
    \caption{An illustration of the mapping between the software hierarchy and the physical GPU hardware hierarchy.}
    \label{fig:execution_model}
    \vspace{-5mm}
\end{figure}

Without loss of generality, our discussion is contextualized within the NVIDIA GPU architecture, primarily focusing on LLM inference scenarios. To analyze performance in this context, we distinguish two tightly related perspectives: the \textit{software} view, which involves how kernels are defined and launched, and the \textit{hardware} view, which describes how kernels are mapped to and executed on the underlying microarchitecture, as illustrated in Figure~\ref{fig:execution_model}.

Kernel execution involves two conceptual stages. The \textbf{compilation stage} produces GPU-executable code in the form of \texttt{SASS} instructions, which are the only representation that NVIDIA SMs can natively execute. In practice, the compilation toolchain may emit a fat binary that contains both architecture-specific \texttt{SASS} \cite{sass_docs} and a virtual instruction set architecture (ISA) (\texttt{PTX} \cite{ptx_docs}); the CUDA driver may further JIT-compile \texttt{PTX} into \texttt{SASS} when necessary. Regardless of whether a kernel originates from Triton, CUDA~C++, or precompiled libraries such as cuBLAS, its final execution always resolves to \texttt{SASS} instructions compatible with the target GPU microarchitecture \cite{nvcc_docs,triton_docs}.

The \textbf{runtime stage} is responsible for launching the compiled kernel. A kernel launch specifies the CUDA execution configuration (Grid and its Cooperative Thread Arrays) \cite{cuda_driver}. The GPU hardware work-distribution logic schedules CTAs onto SMs, dynamically dispatching work in units of CTAs rather than mapping the entire Grid at once \cite{nvidia_cudaguide,cuda_model}. At the SM level, the \texttt{SASS} instructions from compilation are fetched via the instruction cache hierarchy and issued by warp schedulers to the SM’s diverse execution pipelines.

An SM in Ampere \cite{nvidia_ampere_whitepaper} and later architectures is typically partitioned into four SM Sub-partitions (SMSPs) and one Memory I/O (MIO) unit. Each \textbf{SMSP} mainly includes a Warp Scheduler for cycle-by-cycle instruction dispatch, a Register File, and a collection of specialized math pipelines. Notable examples are fixed-latency math pipelines such as FMA (processing most FP32 arithmetic operations like \texttt{FMUL}, \texttt{FADD}) and Tensor (executing MMA instructions like \texttt{HMMA}), as well as variable-latency math pipelines like the XU (for special functions such as base-2 exponential \texttt{MUFU.EX2}). The \textbf{MIO} unit is dedicated to managing data movement. It incorporates on-chip memory caches---the L1 cache and Shared Memory (SMEM)---as well as the Load Store Unit (LSU), which executes memory instructions (e.g., \texttt{LDGSTS}, \texttt{STS}) for accessing global, local, or shared memory \cite{nvidia_forum,nsight_compute_docs}. 

This consistent high-level SM organization across Ampere and later architectures provides a stable micro-architectural abstraction. \textsc{PipeWeave} builds on this foundation to achieve generalizability across kernels and hardware platforms.

\section{Motivation}
\label{sec:motivation}

The rapid co-evolution of LLMs and GPUs necessitates performance modeling tools that are fast, accurate, and generalizable across diverse LLMs, serving frameworks, and hardware architectures. Current approaches, such as cycle-accurate simulation \cite{gpgpusim,accelsim}, analytical modeling \cite{AMALI,GCOM,GPUMech}, and data-driven methods \cite{vidur,simai}, each offer partial solutions, yet none fully meet these requirements. Although cycle-accurate simulators deliver high fidelity, they are slow. Analytical models are labor-intensive and lack generalizability.

In this landscape, data-driven strategies---particularly ``grey-box'' approaches that integrate analytical modeling with ML techniques~\cite{habitat, neusight}---represent a distinct paradigm. These methods aim to deliver rapid predictions with improved generalizability by utilizing strategies like tile-level decomposition and incorporating hardware specifications. However, despite marking notable progress, even these advanced techniques can encounter modeling constraints that limit their accuracy, with prediction errors exceeding 40\% (Section~\ref{sec:eval_kernel}).

\rev{The core issue arises from insufficient microarchitectural fidelity. Specifically, while state-of-the-art simulators like Neusight~\cite{neusight} incorporate workload-level feature engineering (e.g., tile decomposition), they still fall short of true microarchitectural modeling in three key aspects.}

\rev{\textbf{Mismatched Granularity:} Although such methods decompose workloads into thread block tiles, their hardware representation remains coarse-grained. They primarily rely on tile-level descriptors as inputs to an ML model, without explicitly modeling how a tile’s execution translates into specific demands and contention on the SM’s heterogeneous instruction pipelines. For instance, the execution of a single tile inherently involves concurrent activity across multiple hardware units---such as Tensor, FMA, and memory pipelines. A tile-centric abstraction therefore collapses heterogeneous pipeline activity into a single aggregate workload representation, effectively treating the SM as a monolithic black box rather than reflecting its actual execution behavior.}

\rev{\textbf{Inability to Model Fused Kernels:} Such methods model performance primarily at the level of individual deep learning operators in the computation graph (e.g., GEMM, Softmax). This abstraction assumes that kernel execution can be approximated as a composition of standard operators. However, modern high-performance implementations increasingly rely on fused kernels (e.g., FlashAttention~\cite{flashattention}), where multiple operators are tightly integrated into a single GPU kernel. In such kernels, performance is governed by tightly coupled execution patterns introduced by operator fusion, where multiple computations are executed within a shared execution structure and intermediate data is reused across steps. As a result, the effective computation and data movement behavior no longer aligns with the boundaries of individual operators. These execution characteristics cannot be accurately captured through operator-level decomposition, leading to significant modeling limitations.}

\rev{\textbf{Static Wave Modeling:} While current baselines attempt to account for wave quantization (the tail effect of thread block scheduling), they typically rely on a static assumption that tiles within a wave exhibit uniform execution latency. In practice, dynamic workloads---such as causal attention processing variable-length tokens or kernels with early-exit conditions---introduce substantial tile-to-tile latency variation. Without a granular scheduling model, these approaches struggle to capture cross-SM load imbalance and tail effects that frequently degrade real-world performance.
}

\rev{These limitations highlight the need for a new modeling approach that more faithfully reflects GPU execution. An accurate and generalizable model must bridge high-level workload structure with the microarchitectural realities of how kernels execute on modern GPUs. In particular, it should explicitly characterize how workload decomposition and scheduling translate into concrete demands on heterogeneous instruction pipelines, rather than treating execution as an abstract workload description. At the same time, purely analytical modeling is insufficient to capture the complex interactions and resource coupling that arise in real kernels, and such models are often tailored to specific operators and hardware assumptions, making them difficult to generalize as new kernels or GPU architectures emerge without significant re-derivation. Therefore, a hybrid design is needed: one that grounds the model in execution-aware analytical structure while leveraging learning-based components to capture higher-order performance effects. This design philosophy underpins \textsc{PipeWeave}.}

\section{The \textsc{PipeWeave} Design}
\label{sec:design}

Achieving accurate and generalizable GPU performance prediction requires a comprehensive understanding of the intricate interplay between software kernels and underlying hardware architectures. A robust modeling approach must account for both deterministic first-order effects and complex dynamic interactions. Accordingly, we propose \textsc{PipeWeave}, a framework built on a methodology guided by the \textit{dual principles of knowledge and data}.

The \textit{knowledge-driven} component is a hierarchical analytical model that leverages deep domain-specific knowledge of the GPU's parallel execution model to systematically decompose a kernel's complex execution flow. This top-down decomposition progresses from the entire kernel to a set of fundamental tasks, and further into the elemental demands on specific instruction pipelines. This decomposition yields an interpretable feature set for the complementary \textit{data-driven} component: a lightweight MLP designed to capture the complex non-linear interactions and resource contention, which are challenging to characterize analytically. It is this integration of knowledge-driven decomposition and data-driven modeling for higher-order effects that enables \textsc{PipeWeave} to achieve high-fidelity performance predictions.

\begin{figure}[tb]
    \centering
    \hspace*{-0.03\columnwidth}
    \includegraphics[width=0.9\columnwidth]{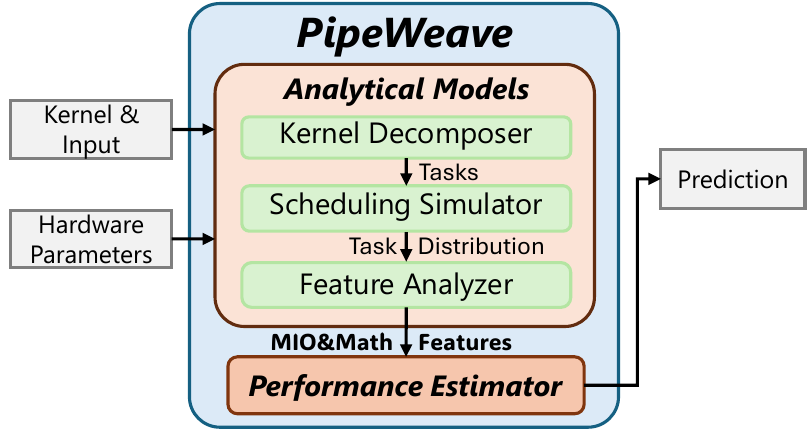}
    \vspace{-2mm}
    \caption{Overview of the \textsc{PipeWeave} modeling framework, detailing the flow from kernel decomposition to the final performance prediction.}
    \label{fig:framework_overview}
    \vspace{-5mm}
\end{figure}

\textsc{PipeWeave} comprises four core modules, as shown in Figure~\ref{fig:framework_overview}: (1)~\textbf{Kernel Decomposer}, which breaks down a kernel's overall execution into a set of fundamental tasks (\S\ref{sec:decomposition}); (2)~\textbf{Scheduling Simulator}, modeling how tasks are assigned to the GPU's SMs and producing the final task distribution (\S\ref{sec:scheduling}); (3)~\textbf{Feature Analyzer}, converting the task distribution into a multi-level feature set that captures instruction pipeline demands and associated theoretical cycles (\S\ref{sec:demand}); and (4)~\textbf{Performance Estimator}, which synthesizes these features into a final prediction using a lightweight MLP to model complex higher-order interactions (\S\ref{sec:prediction}).

This multistage design underpins \textsc{PipeWeave}'s generalizability. The initial two modules ensure \textit{kernel generalizability} by converting any kernel into a uniform task distribution, agnostic to its source. The third module then enables \textit{hardware generalizability} by mapping this distribution to a feature set via a compact vector representing the target GPU's architectural parameters. Once the MLP for a given kernel is trained across various hardware configurations, predicting performance for any new input or GPU---even unseen architectures---becomes highly efficient. The process only involves running fast analytical steps to produce the corresponding feature vector, then performing one forward pass of the MLP, enabling real-time predictions.

While our evaluation is validated on NVIDIA GPU architectures (Table~\ref{tab:dataset_gpus}), the principle of decomposing a kernel into its demands on heterogeneous instruction pipelines is fundamentally general. This can be readily extended to other modern accelerators, such as AMD GPUs.

\subsection{Kernel Decomposer}
\label{sec:decomposition}

To accurately capture the parallel execution of modern GPUs as described in Section \ref{sec:GPU-arc}, \textsc{PipeWeave} decomposes a kernel's workload into a set of smaller \textit{tasks}. This decomposition is central to our approach, as it models the kernel in a manner consistent with GPU parallelism \cite{nvidia_cudaguide}. Although prior studies~\cite{habitat, neusight} have explored partitioning kernels into tiles, they often rely on inferring simplified tiling logic from profiling data. In contrast, \textsc{PipeWeave} emphasizes deterministic analysis of available source code. This enables a more accurate and verifiable decomposition process, capturing complex and diverse task structures in modern kernels.

The precise definition of a task can vary across GPU architectures and kernel implementations. In the \textit{conventional GPU execution model} \cite{nvidia_cudaguide} (e.g., FlashAttention-2 \cite{flashattention2}), a task usually corresponds to a Cooperative Thread Array (CTA), also known as a thread block. A kernel launch generates a grid of CTAs, and the hardware scheduler assigns each CTA to one available SM for the duration of its execution. However, in modern high-performance GPU paradigms such as \textit{persistent kernels} used in patterns like Ping-Pong GEMM \cite{pytorch_ping_pong,cutlass_docs}, this one-to-one mapping no longer holds. Under this execution model, a long-lived CTA stays resident on an SM and serves as a persistent worker. Therefore, the fundamental schedulable unit---our \textit{task}---is not the CTA itself, but a smaller computational packet that the resident CTA fetches from a global work queue.


While the fundamental decomposition methodology is consistent, the specific implementation varies by kernel. To characterize a task's execution properties, our framework identifies \textit{dimensional parameters} ($\mathbf{d}_i$) that define its scope and scale. While these dimensional parameters, and hence the computational workload, are often uniform across all tasks in a kernel (e.g., each GEMM task is typically defined by the same tile dimensions $(tile\_M, tile\_N, tile\_K)$), this is not always the case. A key exception accurs in FlashAttention~\cite{flashattention, flashattention2, flashattention3, flashinfer} when causal masking is applied. Due to the causal constraint, tasks processing  earlier query tokens attend to fewer key/value tokens than those handling later tokens. Thus, even if the nominal task dimensions seem uniform, the actual workload per task can differ significantly.

\begin{table}[tb]
    \centering
    \fontsize{8}{9.6}\selectfont
    \caption{Hardware specifications required by \textsc{PipeWeave}.}   \vspace{-2mm}\label{tab:hardware_params}
    
    \begin{tabular}{ccc}
        \toprule
        \textbf{Parameter} & \textbf{Value Range} & \textbf{Unit} \\ 
        \midrule
        Compute Capability & 8.0 -- 12.0 & - \\
        Number of SMs & 78 -- 188 & - \\
        SM Clock Frequency & 1410 -- 2520 & MHz \\
        \addlinespace
        Tensor Pipe Throughput & 512 -- 4096 & ops/cycle/SM \\
        FMA Pipe Throughput & 64 -- 128 & ops/cycle/SM \\
        XU Pipe Throughput & 16 & ops/cycle/SM \\
        \addlinespace
        Global Memory Bandwidth & 696 -- 4916 & GB/s \\
        L2 Cache Bandwidth & 2430 -- 10400 & GB/s \\
        Shared Memory Bandwidth per SM & 128 & Byte/cycle \\
        Shared Memory Size per SM & 100 -- 228 & KB \\
        Register File Size per SM & 256 & KB \\
        \bottomrule
    \end{tabular}
    \vspace{-4mm}
\end{table}

We formalize the process of deriving these tasks and their parameters through a mapping function $\mathcal{F}$. For a given kernel, $\mathcal{F}$ maps the input parameters $\mathbf{X}$ and the hardware's architectural specifications $\mathbf{S}$ (Table~\ref{tab:hardware_params}) to the full set of tasks $\mathcal{T} = \{\tau_1, \tau_2, \dots, \tau_t\}$:
\begin{equation}
\label{eq:F}
    \{\tau_1, \tau_2, \dots, \tau_t\} = \mathcal{F}(\mathbf{X}, \mathbf{S})
\end{equation}
Each task $\tau_i$ encapsulates a specific part of the kernel's workload, characterized by its dimensional parameter vector $\mathbf{d}_i$. These parameters form the basis for analytically deriving the task's execution properties, such as computational and memory demands, as detailed in the subsequent section (\S\ref{sec:demand}).

The method for deriving the decomposition function $\mathcal{F}$ depends on kernel accessibility. For open-source libraries (e.g., FlashInfer~\cite{flashinfer}), $\mathcal{F}$ is derived by directly extracting the parallelization strategy and thread block mapping logic from the source code. However, this approach does not apply to closed-source libraries such as NVIDIA's cuBLAS\cite{cublas_docs}. To handle such case, we infer the mapping function empirically. For example, to identify the decomposition logic for a cuBLAS GEMM kernel running in BF16 precision, we profile its execution over diverse input matrix dimensions $(M, N, K)$ using tools like the PyTorch Profiler \cite{pytorch_profiler}. By analyzing the profiled data, particularly the correlation between kernel names, the number of CTAs, and input sizes, we reverse-engineer the kernel's implicit task partitioning strategy. This empirical approach enables us to build a surrogate mapping function $\mathcal{F}$ that closely approximates the proprietary decomposition logic.

\subsection{Scheduling Simulator}
\label{sec:scheduling}

A kernel's performance is determined not only by its total workload but also by how that work is allocated across the GPU's parallel resources. After decomposing the kernel into an abstract set of tasks, the next key component of our framework is to simulate the scheduling of these tasks onto SMs. This scheduling analysis converts the task set into a concrete \textit{task distribution}, providing a precise mapping of tasks to specific SMs. This mapping is crucial, as it enables accurate per-SM characterization of the kernel's behavior and helps identify performance bottlenecks resulting from workload imbalance---a critical aspect overlooked in prior studies \cite{roofline, linear_sim,neusight,habitat}. They often rely exclusively on aggregated kernel-level metrics and assume an over-simplified scheduling model where all tasks are handled uniformly. \textsc{PipeWeave} is designed for versatility, supporting the two main scheduling paradigms used in modern GPU applications.

\noindent\textbf{Hardware-Implemented Scheduler.} For conventional kernels, task scheduling is handled by the GPU's hardware scheduler, called the \textit{GigaThread Engine}\cite{li2017cta,nvidia_cudaguide}. Since the exact behavior of this hardware component is not publicly documented, its default scheduling policy is generally inferred from empirical studies to be round-robin (RR)~\cite{hong2009msp, liu2012locality, ji2013dynamic, liu2013greedy, song2016ctascheduling, nvidia2009cuda, jog2016owl, zhang2017tlp, li2017cta}. The policy first assigns each SM at least one task (i.e., a CTA). If an SM still has enough resources (e.g., registers, shared memory, warp-slots, etc.) to support additional tasks, a second assignment round is performed. This rounding-assignment process continues until all SMs are saturated, either due to resource constraints or hardware limits. Afterwards, a new task is assigned to an SM when an existing task finishes and retires from it.

\noindent\textbf{Software-Implemented Scheduler.}
For persistent kernels, the role of hardware scheduler in dispatching CTAs becomes secondary, as each CTA launches only once and remains resident on an SM during execution. Key scheduling logic is handled in software. In this setup, a long-lived CTA repeatedly processes fine-grained work units taken from a global list. In GEMM-like kernels, these units are commonly implemented as \textit{tiles}, which represent the concrete form of our \textit{tasks}. Tile assignment is managed by a tile scheduler\cite{cutlass_docs, colfax_persistent_kernel}, a software component with logic specific to the kernel.

By simulating these scheduling mechanisms, \textsc{PipeWeave} accurately derives a realistic \textit{task distribution}. We formalize the distribution as a partition of the total task set, $\mathcal{T} =\{\tau_1, \tau_2, \dots, \tau_t\}$, across available SMs. This partition comprises sets, $\{\mathcal{T}_1, \mathcal{T}_2, \dots, \mathcal{T}_{N_{SM}}\}$, where $N_{SM}$ denotes the  SM count and each set $\mathcal{T}_j$ contains all tasks assigned to the $j$-th SM. Our scheduling simulator, represented by mapping function $\mathcal{M}$, generates this partition as follows:
\begin{equation}
    \{\mathcal{T}_1, \mathcal{T}_2, \dots, \mathcal{T}_{N_{SM}}\} = \mathcal{M}(\mathcal{T}, \mathbf{S})
\end{equation}
The sets $\{\mathcal{T}_j\}$ form a partition of $\mathcal{T}$, such that $\bigcup_{j=1}^{N_{SM}} \mathcal{T}_j = \mathcal{T}$ and $\mathcal{T}_i \cap \mathcal{T}_j = \emptyset$ for $i \neq j$.

\subsection{Feature Analyzer}
\label{sec:demand}

Feature engineering is conceptually guided by principles from the Roofline performance model~\cite{roofline}. This classic model offers a powerful first-order analysis by determining whether a kernel is bound by the hardware's peak compute throughput or memory bandwidth. However, its predictive accuracy for modern GPUs remains limited. This occurs because its high-level, two-dimensional view of compute and memory fails to capture the intricate resource contention and dynamic interactions that arise when complex modern kernels execute on heterogeneous hardware.

\begin{figure}[tb]
    \centering
    \includegraphics[width=1\columnwidth]{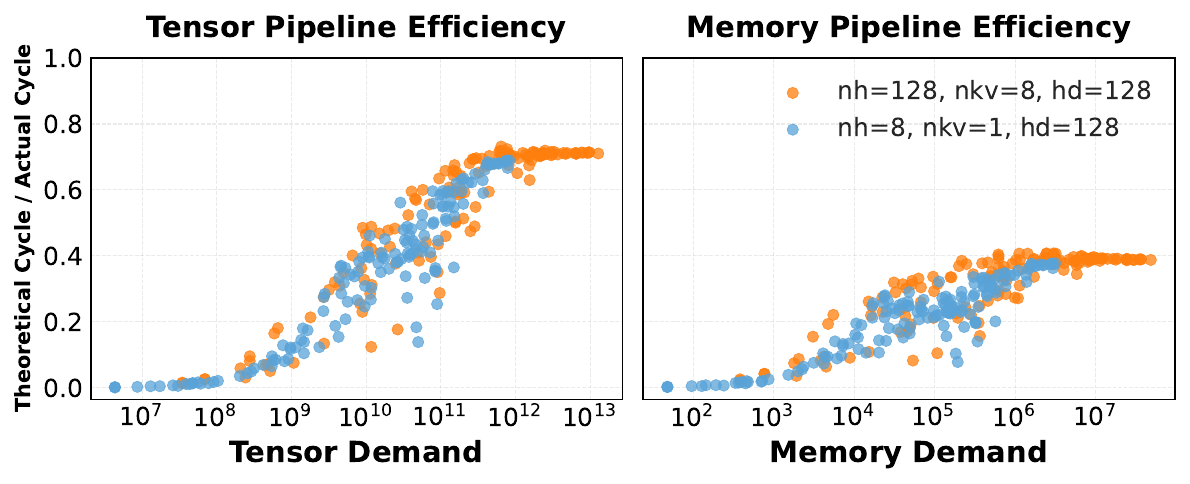}
    \vspace{-7mm}
    \caption{Illustration of the \textsc{PipeWeave} multi-dimensional analysis for FlashAttention-2 on A100. As demand increases, measured performance for two different configurations approaches the theoretical ``roof" and plateaus.}
    \label{fig:attn_roofline}
    \vspace{-4mm}
\end{figure}

To overcome this limitation, \textsc{PipeWeave} expands the Roofline model into a \textit{multi-dimensional analysis}. Instead of a single compute roof and a single memory roof, our model calculates a separate theoretical performance limit for every key instruction pipeline. This necessitates characterizing kernel execution along two fundamental dimensions: (1) \textbf{Demand}, measuring the total workload (e.g., operations or bytes) applied to each pipeline; (2) \textbf{Theoretical Cycles}, obtained from the demand, indicating the ideal execution time if that pipeline alone were the bottleneck. This resembles a particular pipeline's ``roof''. Figure~\ref{fig:attn_roofline} shows a concrete example.
\rev{It plots execution efficiency---the ratio of theoretical cycles to measured latency---against absolute pipeline demand. Unlike the standard roofline, pipelines are decoupled into separate plots, each showing a predictable and independent saturation trend.}

\rev{Moreover, we do not construct rigid analytical models for complex instruction-level concurrency (e.g., the parallel execution of Tensor and FMA pipelines) or architecture-specific mechanisms (e.g., Hopper's Tensor Memory Accelerator (TMA)). Accurately modeling such microarchitectural details would require generation-specific reverse engineering, which undermines cross-generation generalizability and significantly increases modeling complexity.
Instead, \textsc{PipeWeave} adopts a deliberate abstraction strategy. We unify diverse memory access mechanisms—ranging from conventional LSU instructions to advanced asynchronous copies—into generalized memory pipeline demands. By exposing these fundamental pipeline demands as separate raw features, we allow the model to learn their complex and non-linear interactions automatically in the subsequent MLP stage. Empirically, we find that this abstraction is sufficient to capture the dominant performance behaviors across architectures while maintaining strong generalizability.}

The generation of these features follows a bottom-up process across three levels. First, at the \textit{task} level, we characterize the isolated demands of both Math pipelines and MIO pipelines, deriving their corresponding per-task theoretical cycles. Next, these per-task features are aggregated to the \textit{SM} level, creating a detailed profile for each SM and enabling identification of traits for the most heavily utilized SM. Finally, a second aggregation yields a whole-GPU profile containing demand and theoretical cycle metrics for all major pipelines.

\subsubsection{Math Pipelines}

For each task $\tau_i \in \mathcal{T}$, we define its computational demand per math pipeline by the number of executed operations it executes. These pipelines mainly process two operation types: matrix-multiply-accumulate (MMA) operations executed on the Tensor pipeline, and element-wise (EW) operations handled by units like FMA or XU pipelines. Key operations for each math pipeline \cite{nsight_compute_docs,nvidia_forum,nvidia_cudaguide,nvidia_cudaguide1} are outlined in Table~\ref{tab:math_pipeline_ops}.

\begin{table}[tb]
    \centering
    \fontsize{8.5}{10.2}\selectfont
    \caption{Primary operations executed by key math pipelines.}
    \vspace{-2mm}
    \label{tab:math_pipeline_ops}
    \begin{tabularx}{\linewidth}{cX}
        \toprule
        \textbf{Math Pipeline} & \textbf{Primary Operations} \\
        \midrule
        Tensor & MMA instructions across various precisions (e.g., FP8, FP16, BF16). \\
        \midrule 
        FMA    & FP32 floating-point add, multiply, and fused multiply-add. \\
        \midrule 
        XU     & FP32 approximate floating-point special functions (e.g., reciprocal, reciprocal square root, base-2 logarithm, base 2 exponential, sine, cosine). \\
        \bottomrule
    \end{tabularx}
    \vspace{-4mm}
\end{table}

For MMA operations in $\tau_i$, the operation count ($N_{\text{ops,Tensor}}$) is derived directly from the task dimension vector $\mathbf{d}_i$, which includes the tile geometry $\{tile\_M, tile\_N, tile\_K\}$. The total operation count is:
\begin{equation}
    N_{\text{ops,Tensor}} = \alpha \cdot tile\_M \cdot tile\_N \cdot tile\_K
\end{equation}
Here, coefficient $\alpha$ represents the total number of basic multiply-add operations per output element during MMA computations. In a standard GEMM kernel \cite{nvidia_dlperf_gemm, nvidia_cutlass_gemm}, one matrix multiplication gives $\alpha=2$, while a FlashAttention kernel does two sequential matrix multiplications per task \cite{flashattention}, resulting in $\alpha=4$.

For the EW operations in task $\tau_i$, our analysis directly computes the total operations (e.g., $N_{\text{ops,FMA}}$, $N_{\text{ops,XU}}$) for each math pipeline. This entails deriving the aggregate operation counts for specific hardware pipelines (Table~\ref{tab:math_pipeline_ops}) by analyzing the kernel's arithmetic expressions and loop iteration spaces.

Finally, for each pipeline $p$, the theoretical cycles $C_{p}$ needed to execute these operations are determined by dividing the total operation count $N_{\text{ops}, p}$ by its corresponding throughput $Th_{p}$, a parameter from hardware specification $\mathbf{S}$:
\begin{equation}
    C_{p} = \frac{N_{\text{ops}, p}}{Th_{p}}
    \label{eq:ops_to_cycles}
\end{equation}

After obtaining per-task demand features, we use a bottom-up approach to aggregate task distributions $\{\mathcal{T}_1, \mathcal{T}_2, \\\dots, \mathcal{T}_{N_{SM}}\}$ into SM-level and GPU-level features. Starting at the SM level, for each pipeline $p$, we combine the demands of all tasks assigned to SM$_j$ to compute total per-SM operations $N_{\text{ops}, p}^{\text{SM}_j}$ and theoretical cycles $C_{p}^{\text{SM}_j}$. These per-SM values are summed to obtain overall GPU operations $N_{\text{ops}, p}^{\text{GPU}}$. Corresponding GPU-level theoretical cycles are derived from this total workload and the combined throughput of pipeline $p$:
\begin{equation}
    C_{p}^{\text{GPU}} = \frac{N_{\text{ops}, p}^{\text{GPU}}}{N_{SM} \cdot Th_{p}}
\end{equation}

\subsubsection{MIO pipelines}

For MIO pipelines, we measure total demand in bytes at three levels. First, for each task $\tau_i$, we calculate the total \textit{per-task} memory demand $B_i$ by summing all data it \texttt{load}s from the memory hierarchy. This approach is taken because \texttt{load}s are often on the critical execute path in most kernels. A data stall directly affects consumer latency (math pipelines) \cite{nsight_compute_docs}. Using the task distribution, these per-task values are summed for tasks in set $\mathcal{T}_j$ to get the \textit{per-SM} memory demand $B^{\text{SM}_j}$. Finally, summing all per-SM values $B^{\text{SM}_j}$ gives the \textit{global} memory demand $B^{\text{GPU}}$.

From these aggregated byte counts, we derive several theoretical cycle features. The theoretical cycles $C_{\text{mem}}$ is calculated by dividing total bytes at a given level by a specific memory subsystem's theoretical bandwidth, expressed as $C_{\text{mem}} = B / BW_{\text{mem}}$. At GPU-level, we apply this formula with $B^{\text{GPU}}$, using L2 Cache and Global Memory bandwidths. At SM-level, $B^{\text{SM}_j}$ is used along with per-SM bandwidths for Shared Memory, L2 Cache, and Global Memory.

\subsection{Performance Estimator}
\label{sec:prediction}

The final component of \textsc{PipeWeave} is a lightweight machine learning model that predicts the overall kernel execution duration. The MLP uses a single feature vector as input, which is the concatenation of all analytical features from earlier stages (Section~\ref{sec:demand}). This vector includes features (Table~\ref{tab:mlp_feature_vector}) from the MIO pipeline, plus features from one or more Math pipelines based on the kernel's specific operations.

We adopt a per-kernel modeling approach, training a separate MLP for each kernel category. Each MLP's training dataset is built by profiling the corresponding kernel's execution across various GPU architectures and input parameters. For every sample, we record the actual execution latency on physical hardware as ground-truth.

\begin{table}[tb]
    \fontsize{8.5}{10.2}\selectfont
    \centering
    \caption{The analytical feature vector provided as input to the MLP.}
    \vspace{-2mm}
    \label{tab:mlp_feature_vector}
    \begin{tabularx}{\linewidth}{llX}
        \toprule
        \textbf{Pipeline} & \textbf{Granularity} & \textbf{Features} \\
        \midrule
        \multirow{4}{*}{Math} & \multirow{2}{*}{GPU} & Total Operations \\
                              &                              & Total Theoretical Cycles \\
                              \cmidrule(l){2-3} 
                              & \multirow{2}{*}{SM}  & Max SM Operations \\
                              &                              & Max SM Theoretical Cycles \\
        \midrule
        \multirow{5}{*}{MIO}  & \multirow{2}{*}{GPU} & Total Memory Demand \\
                              &                              & Theoretical Cycles (Global, L2) \\
                              \cmidrule(l){2-3}
                              & \multirow{2}{*}{SM}  & Max SM Memory Demand \\
                              &                              & Theoretical Cycles (Global, L2, Shared) \\
        \bottomrule
    \end{tabularx}
    \vspace{-4mm}
\end{table}

\section{Implementation Details}
\subsection{Analytical Models}

To ensure our performance model accurately reflects real-world LLM inference workloads, we chose a representative set of critical kernels directly from the backends of popular high-performance serving frameworks, such as SGLang~\cite{sglang} and vLLM~\cite{vllm}. The key characteristics of these kernels are summarized in Table~\ref{tab:kernel_selection}. Note that for categories like GEMM and Attention, multiple implementations often exist. For cuBLAS GEMM kernels, we observe that specific implementations vary across hardware architectures. For FlashInfer Attention kernels, our analysis includes both FlashAttention-2 (FA2) and FlashAttention-3 (FA3) variants, covering implementations for paged and ragged KV cache layouts \cite{flashinfer_kv_layout}. 

\begin{table}[tb]
    \vspace{-3mm}
    \fontsize{6.5}{7}\selectfont
    \centering
    \caption{Key characteristics of the kernels selected.}
    \vspace{-2mm}
    \label{tab:kernel_selection}
    \begin{tabularx}{\linewidth}{Xlllll} 
        \toprule
        \textbf{Category} & \textbf{Source} & \textbf{Language} & \textbf{Precision} & \textbf{Scheduler} & \textbf{Math Pipe} \\ 
        \midrule
        GEMM & cuBLAS & Pre-compiled & BF16/FP16 & HW/SW & Tensor \\ 
        \addlinespace
        Scaled MM & vLLM & CUDA C++ & FP8 & HW/SW & Tensor \\ 
        \addlinespace
        Attention & FlashInfer & CUDA C++ & BF16/FP16 & HW/SW & Tensor, XU \\ 
        \addlinespace
        RMSNorm & FlashInfer & CUDA C++ & FP32 & HW & FMA, XU \\
        \addlinespace
        SiLU\&Mul & FlashInfer & CUDA C++ & FP32 & HW & FMA, XU \\
        \addlinespace
        Fused MoE & SGLang & Triton & BF16/FP16 & HW & Tensor \\ 
        \bottomrule
    \end{tabularx}
    \vspace{-4mm}
\end{table}

For each kernel category, the implementation of the \textbf{Kernel Decomposer} is concise, requiring just 10-50 lines of code. Except for cuBLAS GEMM whose decomposition is taken directly from profiling data, the mapping function $\mathcal{F}$ in Equation (\ref{eq:F}) for other kernels is drawn from their source code. Because cuBLAS GEMM is closed-source and its implementation differs across hardware architectures, its decomposition behavior is unknown on new GPUs. Therefore, for unseen GPUs lacking profiling data on closed-source kernels, we use decomposition logic from the most architecturally similar GPUs available in our profiling dataset. 

Following kernel decomposition, the \textbf{Scheduling Simulator} allocates tasks across SMs. For the majority of kernels analyzed (Table~\ref{tab:kernel_selection}), which utilize the hardware-based scheduler, we simulate the widely inferred RR policy as described in Section~\ref{sec:scheduling}. For cuBLAS GEMM and FlashInfer FA3 kernels~\cite{flashinfer} on the Hopper architecture, both using persistent kernel designs, we model their respective software-based schedulers. Taking FlashInfer FA3 as an example, we accurately replicated its MinHeap-based scheduler logic in our simulator with around 40 code lines.

The \textbf{Feature Analyzer} converts the task distribution into a comprehensive feature set. For math pipelines, our implementation focuses on three types of instruction pipelines most critical to LLM workloads: the Tensor, FMA, and XU pipelines. We found that these three together cover most computational demands in the target kernels. Other pipelines, such as ALU handling logic operations \cite{nsight_compute_docs}, were left out due to their generally low utilization in the kernels and the difficulty in analytically counting their operations.

\subsection{Dataset Construction}
\label{sec:dataset}

To train and evaluate \textsc{PipeWeave}, we built a comprehensive dataset by profiling selected kernels (Table~\ref{tab:kernel_selection}) across various NVIDIA GPU architectures. The dataset covers 11 different GPU models \cite{nvidia_ampere_whitepaper,nvidia_ada_whitepaper,nvidia_hopper_whitepaper,nvidia_blackwell_whitepaper}, representing multiple architectures and market segments. As shown in Table~\ref{tab:dataset_gpus}, these were split into two groups: the first group was used for training, while the second group was reserved solely for testing to assess \textsc{PipeWeave}'s generalizability to unseen hardware.

\begin{table}[tb]
    \centering
    \fontsize{6.6}{7}\selectfont 
    \caption{Key specifications of the evaluated NVIDIA GPUs.}
    \vspace{-2mm}
    \label{tab:dataset_gpus}
    \begin{tabularx}{\linewidth}{lXrrrr}
        \toprule
        \textbf{GPU} & \textbf{Architecture} & \textbf{SMs} & \textbf{Mem BW} & \textbf{Tensor BF16} & \textbf{Freq} \\
        & & & (GB/s) & (ops/clk/SM) & (MHz) \\ 
        \midrule
        A40            & Ampere     & 84   & 696   & 1024 & 1740 \\ %
        A100           & Ampere     & 108  & 2039  & 2048 & 1410 \\ %
        RTX 6000 Ada   & Ada        & 142  & 960   & 1024 & 2505 \\ %
        L20            & Ada        & 92   & 864   & 516  & 2520 \\ %
        H20            & Hopper     & 78   & 4023  & 1024 & 1830 \\ %
        H800           & Hopper     & 132  & 3352  & 4096 & 1830 \\ %
        \midrule 
        RTX A6000      & Ampere     & 84   & 768   & 1024 & 1800 \\ %
        L40            & Ada        & 142  & 864   & 512  & 2490 \\ %
        H100           & Hopper     & 132  & 3352  & 4096 & 1830 \\ %
        H200           & Hopper     & 132  & 4917  & 4096 & 1830 \\ %
        RTX PRO 6000 S & Blackwell  & 188  & 1792  & 1024 & 2340 \\ %
        \bottomrule
    \end{tabularx}
    \vspace{-4mm}
\end{table}

Profiling was performed in a consistent software environment using PyTorch 2.8.0, CUDA Toolkit 12.8, FlashInfer 0.4.1, SGLang 0.5.4, vllm 0.11.0, and Triton 3.4.0. For each combination of kernel, input parameters, serving framework, and GPU hardware, we measured execution latency with the PyTorch Profiler. We conducted 5 warm-up runs followed by 10 measurement runs, using their average as the ground-truth. The profiling dataset includes 6 key kernels serving as core computational backend for vLLM~\cite{vllm} and SGLang~\cite{sglang}:

\begin{itemize}[leftmargin=*]
  \item \textbf{Attention}: 104,958 samples (71,969 training and 32,989 test). $bs \in [1, 16]$, $nh \in [2, 128]$, $nkv \in [1, 8]$, $hd \in \{64,
128\}$, $qlen \in [1, 20097]$, $kvlen \in [4, 20481]$. The Query and KV lengths vary randomly within each batch to simulate realistic variable-length
sequence patterns.

  \item \textbf{GEMM}: 613,263 samples (494,463 training and 118,800 test). $M \in [2, 131072]$, $N \in [384, 152064]$, $K \in [256, 53248]$.

  \item \textbf{RMSNorm}: 65,036 samples (44,592 training and 20,444 test). $seq \in [2, 131072]$, $dim \in [128, 16384]$.

  \item \textbf{SiLU\&Mul}: 104,834 samples (71,868 training and 32,966 test). $seq \in [2, 131072]$, $dim \in [768, 106496]$.

  \item \textbf{Scaled MM}: 25,228 samples (16,818 training and 8410 test). $M \in [2, 131072]$, $N \in [384, 8192]$, $K \in [256, 8192]$.
  
  \item \textbf{Fused MoE}: 33,264 samples. $M \in [2, 8192]$,
  $E \in [8, 128]$, $topk \in [2, 8]$, $H \in [1024, 4096]$, $N \in [512, 3072]$. This kernel is used as a detailed case study for our optimization approach in Section~\ref{sec:optimization_guidance}.
\end{itemize}

\subsection{MLP Model Training}
\label{sec:impl_training}

As outlined in Section~\ref{sec:prediction}, a separate MLP is trained for each kernel type using derived analytical features. The MLP has a shallow architecture with 3 hidden layers (256, 128, and 64 units), employing ReLU activations followed by Batch Normalization and Dropout (rate 0.1) for regularization. The output layer utilizes a Sigmoid activation to limit predictions to the range $[0, 1]$, representing the kernel's \textit{execution efficiency} (defined as the ratio of theoretical execution time to actual latency). The final latency prediction is obtained by dividing the theoretical execution time by this estimated efficiency.

Training uses the dataset described in Section~\ref{sec:dataset}. The AdamW optimizer~\cite{adamw} is applied with a 0.001 initial learning rate and weight decay. Mean Absolute Percentage Error (MAPE) serves as the loss function, minimizing relative prediction error. Early stopping is employed to prevent overfitting by monitoring validation loss. 

\subsection{End-to-end Performance Prediction}

\label{sec:impl_e2e}

Beyond predicting single kernel performance, we validate our framework's accuracy in modeling end-to-end LLM inference latency. We built a \textbf{Workload Generator} based on the model definitions and kernel invocation logic from both SGLang~\cite{sglang} and vLLM~\cite{vllm}. Given a model configuration and input parameters, this generator creates a sequence of kernel invocations that represents a real inference scenario. Following prior work~\cite{habitat,neusight,daydream}, we assume sequential kernel execution without overlap. For each kernel in the sequence, we use \textsc{PipeWeave} to predict its runtime based on type and input dimensions. The total end-to-end latency for single-GPU inference is calculated by summing all predicted kernel durations.

Predicting end-to-end performance in distributed settings requires modeling both computational kernels and communication kernels required for multi-GPU parallelism \cite{vidur,neusight,llmcompass,simai}. Depending on the employed parallelism, this includes kernels such as \textit{All-Reduce} for Tensor Parallelism (TP) or \textit{Send/Recv} primitives for Pipeline Parallelism (PP). To model these communication kernels, we use a simplified method. We profile their performance across different network topologies and communication volumes to build a baseline performance database. Using this data, we apply a data-driven regression technique (e.g., Random Forest) to estimate communication kernel latency. This prediction is then combined with computational kernel estimates to forecast the total end-to-end latency for distributed inference.

\section{Evaluation}
\subsection{Baselines}
\label{sec:baselines}

To comprehensively evaluate \textsc{PipeWeave}, we conduct our primary evaluation by comparing its prediction accuracy against \rev{four} main baselines: (1) the classic analytical Roofline model~\cite{roofline}; (2) a Linear regression-based model~\cite{linear_sim}; \rev{(3) Habitat~\cite{habitat}}; and (4) Neusight~\cite{neusight}, a state-of-the-art data-driven method. 
To ensure a fair comparison among these primary baselines, we adjusted them to incorporate our analytical components. The Linear model, following the approach in the original paper~\cite{linear_sim}, was trained using two main features from our Feature Analyzer (Section~\ref{sec:demand}): theoretical cycles for aggregating compute and memory demand. Similarly, we supplied Habitat and Neusight with the exact task definitions from our Kernel Decomposer (Section~\ref{sec:decomposition}).

\rev{Furthermore, to highlight the advantages of our analytical–ML hybrid design in both prediction accuracy and simulation efficiency, we introduce a secondary set of baselines representing highly detailed modeling paradigms: AMALI~\cite{AMALI}, an instruction-trace-based analytical model, and LLMCompass~\cite{llmcompass}, a hybrid framework that integrates analytical models and cycle-accurate systolic array modeling. Since these detailed simulators provide limited support for diverse modern kernels and incur substantial runtime overhead for end-to-end LLM workloads, we restrict this comparison to standalone GEMMs.}



\subsection{Validation of Analytical Components}
\label{sec:eval_analytical}

We first validate \textsc{PipeWeave}'s core analytical components: \textit{Kernel Decomposer}, \textit{Scheduling Simulator}, and \textit{Feature Analyzer}. This step is essential since these parts work in sequence. Any error may propagate and reduce the final feature quality.

We start by verifying the correctness of the \textit{Kernel Decomposer}. Specifically, we compare the number of CTAs from our decomposition process with the ground-truth configurations in the dataset across multiple kernels. The results are fully consistent, confirming decomposition accuracy.

Next, we assess the accuracy of the \textit{Scheduling Simulator} and \textit{Feature Analyzer}. Our method compares analytically derived math pipeline operation counts, both total (kernel-wide) and per-SM maximum operations, against ground-truth measurements from the NVIDIA Nsight Compute (NCU) tool~\cite{nsight_compute_docs}. Due to high profiling overhead and restricted hardware access, we perform this validation on two flagship devices: A100 and H100. The evaluation covers four key kernel implementations: cuBLAS GEMM (gemm8 on A100 and gemm9 on H100), FA2, and FA3. \rev{Each includes about 500 test samples randomly sampled from the workload configuration ranges defined in Section~\ref{sec:dataset}}.
As shown in Table~\ref{tab:analytical_validation_results}, our model achieves a maximum error of \textbf{0.5\%} for total operations and \textbf{6.3\%} for the maximum per-SM operations. \rev{The higher error for FA2 (6.34\%) relative to FA3 (0.45\%) is mainly due to their different scheduling mechanisms: FA3 uses a persistent-kernel design with deterministic task scheduling that can be explicitly simulated, whereas FA2 relies on dynamic hardware scheduling, which introduces additional uncertainty in predicting peak per-SM workload.}

\begin{table}[t]
    \centering
    \fontsize{8}{9.6}\selectfont
    \caption{MAPE (\%) of \textsc{PipeWeave}'s analytical operation counts}
    \vspace{-2mm}
    \label{tab:analytical_validation_results}
    \begin{tabular}{lccccc} 
        \toprule
        \textbf{Metric} & \textbf{gemm8} & \textbf{gemm9} & \textbf{FA2} & \textbf{FA3} \\
        \midrule
        Max SM Ops (\%) & 0.07 & 0.04 & 6.34 & 0.45 \\
        Total Ops (\%)  & 0.01 & 0.14 & 0.50 & 0.00  \\
        \bottomrule
    \end{tabular}
    \vspace{-4mm}
\end{table}

\begin{figure}[b]
    \vspace{-4mm}
    \centering
    \includegraphics[width=0.9\columnwidth]{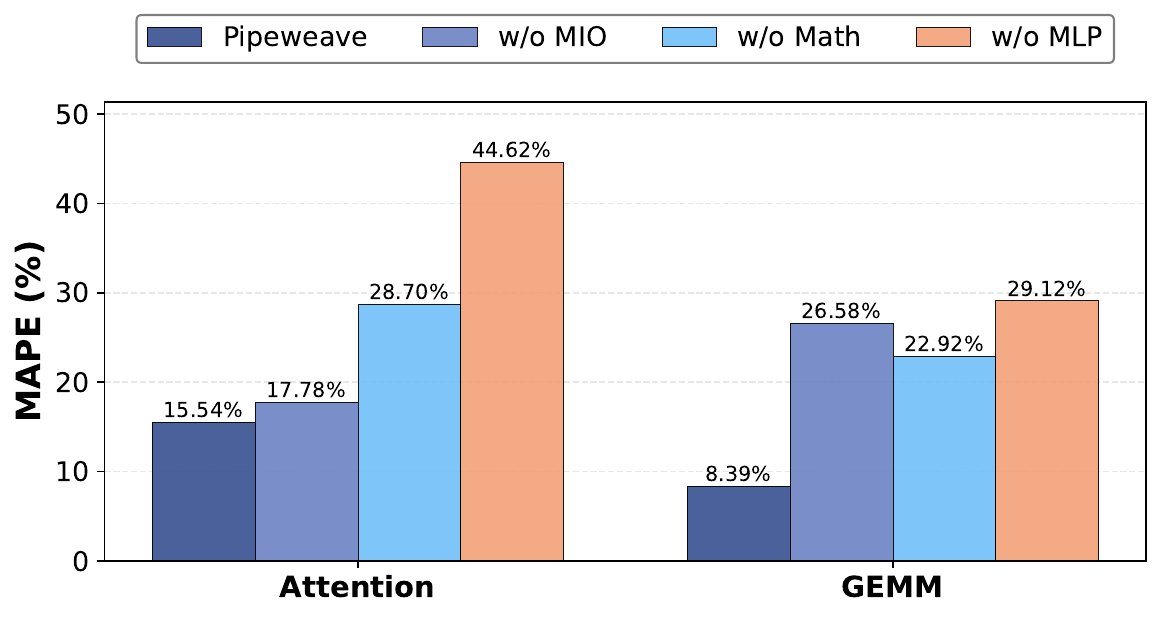}
    \vspace{-4mm}
    \caption{Ablation study on the impact of MIO and Math Pipeline features for GEMM and Attention kernels.}
    \label{fig:feature_ablation}
\end{figure}

Finally, we conduct an ablation study on the GEMM and Attention kernels using their full datasets (Section~\ref{sec:dataset}) to highlight the contribution of our core components. We compare the full \textsc{PipeWeave} model against three ablated variants: (1) w/o MIO (without MIO features), (2) w/o Math, and (3) w/o MLP (replacing MLP with a Roofline-based predictor). As shown in Figure~\ref{fig:feature_ablation}, each component is crucial for accurate performance. For the Attention kernel, the full model achieves \textbf{1.1$\times$}, \textbf{1.8$\times$} and \textbf{2.9$\times$} higher accuracy than w/o MIO, w/o Math, and w/o MLP respectively. The effect is stronger for GEMM, where our full model improves accuracy by \textbf{3.2$\times$} (w/o MIO), \textbf{2.7$\times$} (w/o Math), and \textbf{3.5$\times$} (w/o MLP), respectively.
\rev{While both kernels benefit significantly from our modeling framework, the final prediction error for Attention kernels (15.54\%) remains higher than that of GEMM kernels (8.39\%). As previously shown in Table~\ref{tab:analytical_validation_results}, this gap is not caused by inaccuracies in the analytical operation counts, which remain comparably low for both kernels. Instead, it arises from the inherently uneven workload distribution and dynamic execution characteristics of Attention mechanisms. Unlike GEMM, where tasks are defined by uniform dimensional parameters across tiles, Attention workloads exhibit substantial variance. This variance primarily results from causal masking---where tasks processing earlier query tokens attend to fewer key/value tokens than those handling later tokens---as well as randomly varying sequence lengths within a batch. In addition, Attention introduces more complex memory behavior and heterogeneous execution phases with different compute--memory characteristics, which further increase runtime variability. These factors make execution latency more sensitive to hardware scheduling dynamics and lead to larger inter-block completion variance. Consequently, Attention latency is inherently more difficult for the MLP to model than the stable and uniform execution patterns observed in GEMM workloads.}

\begin{figure*}[t]
    \centering
    \includegraphics[width=\textwidth]{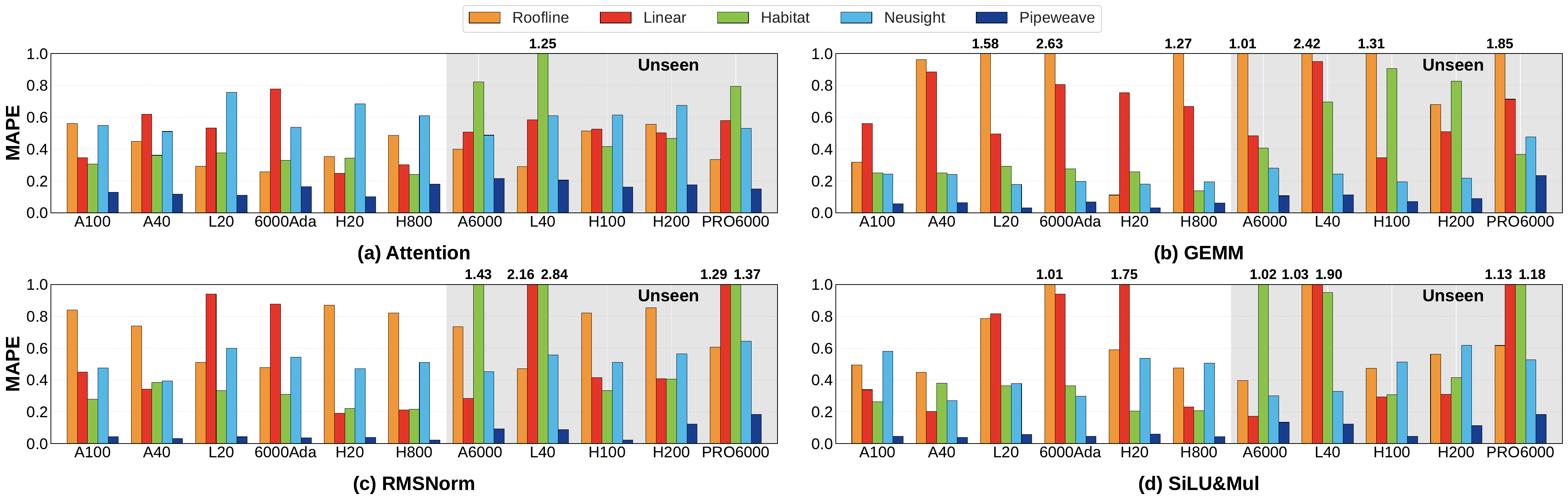}
    \vspace{-8mm}
    \caption{Kernel-level prediction accuracy (MAPE) of \textsc{PipeWeave} and baseline models. Unseen hardware platforms are identified by a grey background.}
    \label{fig:kernel_level_accuracy}
    \vspace{-3mm}
\end{figure*}

\begin{figure*}[t]
    \centering
    \includegraphics[width=\textwidth]{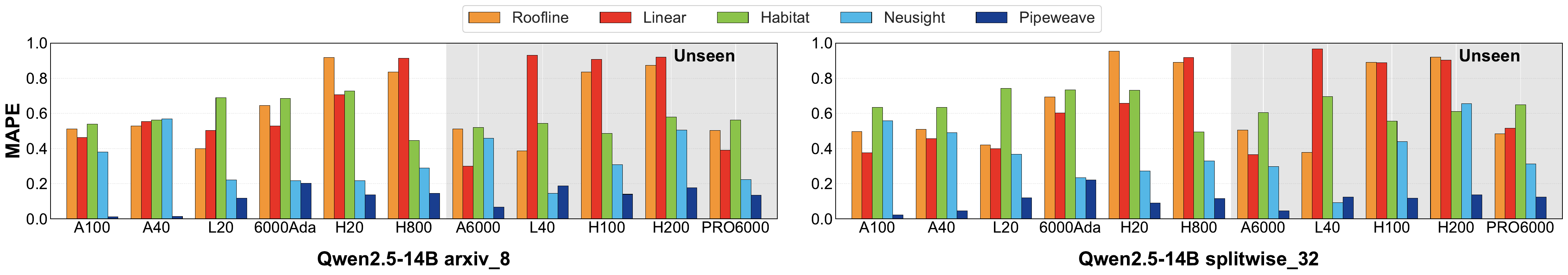}
    \vspace{-8mm}
    \caption{\rev{End-to-end inference prediction accuracy (MAPE) of \textsc{PipeWeave} and baseline models for single-GPU Qwen2.5-14B inference using SGLang. Unseen hardware platforms are identified by a grey background.}}
    \label{fig:e2e_accuracy}
    \vspace{-4mm}
\end{figure*}

\begin{table}[t]
    \centering
    \caption{\rev{Prediction Error on Seen and Unseen GPUs.}}
    \label{tab:mape_results}
    \renewcommand{\arraystretch}{1.2}
    \setlength{\tabcolsep}{6pt} 
    \begin{tabular}{@{}lccccc@{}}
    \toprule
    \textbf{Hardware} & \textbf{Roofline} & \textbf{Linear} & \textbf{Habitat} & \textbf{Neusight} & \textbf{\textsc{PipeWeave}} \\ \midrule
    \textbf{Seen}   & 72.22\% & 59.50\% & 28.92\% & 43.49\% & \textbf{6.77\%} \\
    \textbf{Unseen} & 79.61\% & 70.28\% & 85.96\% & 46.70\% & \textbf{13.14\%} \\ \bottomrule
    \end{tabular}
\end{table}

\subsection{Kernel-Level Prediction Accuracy}
\label{sec:eval_kernel}
We evaluate \textsc{PipeWeave} on a dataset of around \textbf{1M} samples from 11 different GPUs (Section~\ref{sec:dataset}). This dataset includes fundamental kernels from modern inference frameworks such as vLLM \cite{vllm} and SGLang \cite{sglang}), covering FP32, BF16/FP16, and FP8 precisions. \textsc{PipeWeave} achieves state-of-the-art prediction accuracy and significantly surpasses prior work. On seen hardware, it attains an average MAPE of 6.0\%, outperforming the next-best Neusight at 42.6\%. On unseen hardware, our framework demonstrates superior generalization with an average MAPE of 11.5\%-a \textbf{3.9$\times$} improvement compared to Neusight (45.1\%).

Figure~\ref{fig:kernel_level_accuracy} shows the prediction accuracy (MAPE) for four typical kernels in BF16 LLM inference scenarios. \rev{Correspondingly, Table~\ref{tab:mape_results} summarizes the average MAPE across these four kernels on both seen and unseen hardware.} Errors for Linear and Roofline models are significantly higher than \textsc{PipeWeave} on both seen and unseen hardware, with peak MAPEs reaching 215.6\% and 263.5\%, respectively. Although the SOTA baseline, Neusight, outperforms other baselines, its highest error of 75.7\% remains substantially above \textsc{PipeWeave}'s 23.4\%.
\rev{Furthermore, the prediction errors of analytical approaches, namely the Linear and Roofline models, are highly hardware-dependent. For instance, Figure~\ref{fig:kernel_level_accuracy}(b) highlights a stark contrast in the Roofline model's MAPE for GEMM kernels between the H20 (11\%) and H800 (127\%). This difference arises from the distinct compute-to-memory ratios of the two GPUs. Specifically, while the H20 retains approximately 120\% of the H800's memory bandwidth, its peak compute capability is restricted to roughly 15\% of the H800's. Under this extremely low compute-to-memory ratio, the compute units on the H20 are easily saturated. The abundant memory bandwidth ensures that execution pipelines are constantly fed, allowing GEMMs to sustain throughput very close to the theoretical peak; thus, the Roofline estimate remains accurate. Conversely, the H800 features a massive compute capacity that is exceedingly difficult to fully saturate in most practical scenarios, as reaching the theoretical peak requires near-perfect instruction-level concurrency and uninterrupted data delivery. In practice, inevitable microarchitectural frictions prevent kernels from approaching this idealized Roofline peak, leading to significant overestimation. Unlike such models, \textsc{PipeWeave}'s MLP naturally learns these hardware-specific inefficiencies, thereby achieving significantly lower prediction errors.}

Besides the four kernels common in BF16 LLM inference scenarios, we also trained and tested the scaled mm kernel (block-wise quantization) for FP8 inference on the Hopper architecture, achieving high prediction accuracy. On seen hardware (H20, H800), \textsc{PipeWeave}'s MAPE was 1.9\% and 4.1\%, while on unseen hardware (H100, H200), MAPE was 4.2\% and 5.2\%. This highlights the framework's adaptability to FP8 precision kernels, achieving average accuracy gains of \textbf{10.8$\times$}, \textbf{9.5$\times$}, \textbf{5.5$\times$}, and \textbf{7.8$\times$} over Roofline, Linear, Habitat, and Neusight.

\begin{figure}[t]
    \centering
    \includegraphics[width=\columnwidth]{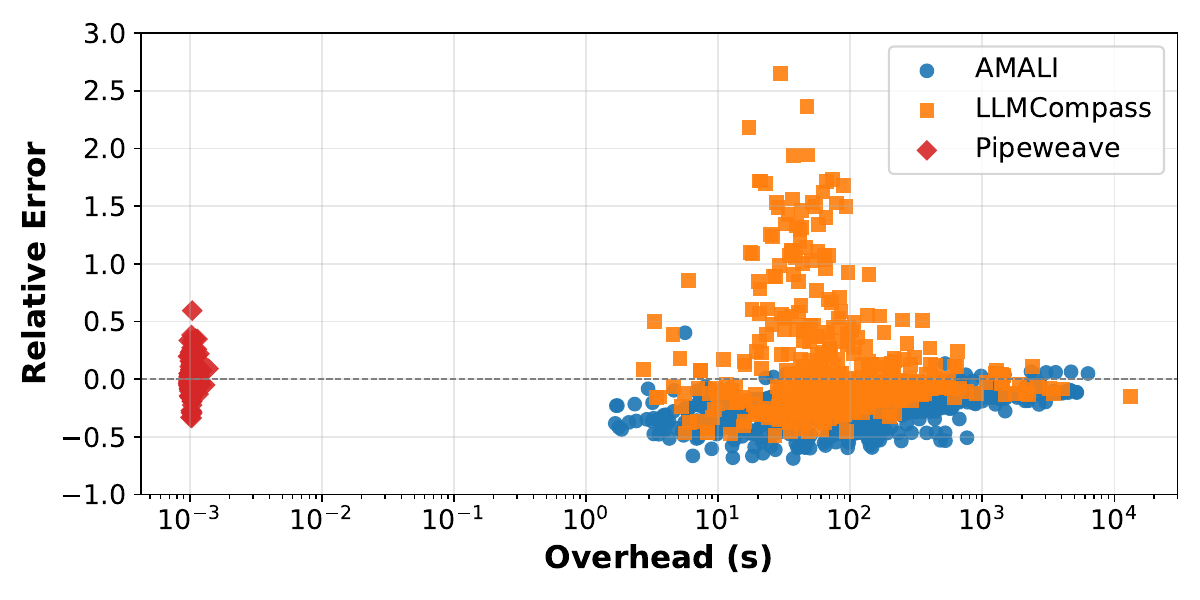}
    \caption{\rev{Comparison of simulation overhead versus relative prediction error for GEMM workloads on the A100 GPU.}}
    \label{fig:gemm_overhead_error}
\end{figure}

\rev{Finally, to evaluate \textsc{PipeWeave}'s prediction accuracy and simulation efficiency, we conduct a targeted comparison with AMALI and LLMCompass on an A100 GPU. As outlined in our baseline methodology (Section~\ref{sec:baselines}), this comparison is restricted to GEMMs due to the high computational overhead of these detailed simulators. Using 540 distinct GEMM samples with varying dimensions randomly drawn from our dataset (Section~\ref{sec:dataset}), we measure prediction error and per-GEMM simulation overhead. Figure~\ref{fig:gemm_overhead_error} shows the comparison results, where prediction error is reported as signed relative error to capture both over- and under-estimation. Overall, \textsc{PipeWeave} achieves substantially lower simulation overhead while maintaining higher prediction accuracy. On average, it obtains a MAPE of \textbf{6.4\%}, compared with \textbf{28.3\%} for AMALI and \textbf{29.7\%} for LLMCompass, while reducing prediction time by \textbf{3 to 7 orders of magnitude}. These results indicate that the grey-box design---combining pipeline-demand analytical modeling with ML---can effectively capture dominant performance factors without requiring expensive low-level simulation.}

\subsection{E2E Inference Accuracy}
\label{sec:eval_e2e}

Beyond kernel-level validation, we assess \textsc{PipeWeave}'s end-to-end predictive accuracy by comparing its simulations with actual serving latencies from SGLang~\cite{sglang} and vLLM~\cite{vllm}. Following prior work~\cite{vidur}, we use two representative datasets Arxiv Summarization~\cite{cohan2018discourse} and Splitwise~\cite{splitwise}—and test three typical LLMs (Qwen2.5-14B, Qwen3-32B, and Llama3.1-70B) in both single-GPU (TP=1) and distributed (TP, PP) inference settings. 

Workloads for these datasets are generated by randomly sampling requests to create batches of varying sizes, such as \texttt{arxiv\_8} and \texttt{splitwise\_64}. The \texttt{arxiv\_*} (where \texttt{*} denotes the batch size) workloads have an average input length of 2{,}630 tokens, while the \texttt{splitwise\_*} workloads average 982 tokens. Output lengths vary from 5 to 4,056 tokens.

\begin{table*}[t]
    \centering
    \fontsize{7}{8}\selectfont
    \caption{\rev{End-to-end performance prediction MAPE (\%) of \textsc{PipeWeave} and baselines for multi-GPU inference using SGLang and vLLM.}}
    \vspace{-2mm}
    \label{tab:e2e_distributed_results}
    \begin{tabular*}{\textwidth}{l @{\extracolsep{\fill}} l @{\extracolsep{\fill}} l @{\extracolsep{\fill}} l @{\extracolsep{\fill}} r @{\extracolsep{\fill}} r @{\extracolsep{\fill}} r @{\extracolsep{\fill}} r @{\extracolsep{\fill}} r}
        \toprule
        \textbf{Framework} & \textbf{Model} & \textbf{Dataset} & \textbf{Hardware} & \textbf{Roofline} & \textbf{Linear} & \textbf{Habitat} & \textbf{Neusight} & \textbf{\textsc{PipeWeave}} \\
        \midrule
        \multirow{8}{*}{SGLang} & \multirow{8}{*}{Qwen3-32B (TP=2)} & 
            \multirow{4}{*}{\texttt{arxiv\_12}} & A100 & 48.6 & 42.7 & 47.3 & 45.0 & \textbf{2.4} \\
            & & & 6000Ada & 59.2 & 43.3 & 44.9 & 30.4 & \textbf{9.1} \\
            & & & H100 & 73.5 & 77.1 & 34.9 & 31.1 & \textbf{7.5} \\
            & & & PRO6000 & 46.5 & 15.2 & 39.6 & 56.6 & \textbf{9.3} \\
        
        \cmidrule(){3-9}
        & & \multirow{4}{*}{\texttt{splitwise\_48}} & A100 & 49.0 & 44.6 & 35.5 & 45.9 & \textbf{3.9} \\
            & & & 6000Ada & 53.2 & 51.5 & 35.1 & 38.8 & \textbf{7.9} \\
            & & & H100 & 62.4 & 49.6 & 33.3 & 60.2 & \textbf{16.5} \\
            & & & PRO6000 & 47.1 & 29.8 & 36.9 & 18.5 & \textbf{10.9} \\
        
        \midrule 
        \multirow{4}{*}{SGLang} & \multirow{4}{*}{Llama3.1-70B (TP=4)} & 
            \multirow{2}{*}{\texttt{arxiv\_16}} & A100 & 45.2 & 30.3 & 50.1 & 76.5 & \textbf{2.6} \\
            & & & H100 & 78.6 & 69.4 & 45.6 & 34.5 & \textbf{5.4} \\
        
        \cmidrule(){3-9}
        & & \multirow{2}{*}{\texttt{splitwise\_64}} & A100 & 46.0 & 26.2 & 57.5 & 55.6 & \textbf{2.1} \\
            & & & H100 & 82.2 & 64.8 & 56.1 & 47.2 & \textbf{7.0} \\

        \midrule 
        \multirow{4}{*}{SGLang} & \multirow{4}{*}{Llama3.1-70B (TP=8)} & 
            \multirow{2}{*}{\texttt{arxiv\_16}} & H20 & 90.1 & 70.5 & 54.4 & 27.1 & \textbf{4.0} \\
        & & & H800 & 66.7 & 46.7 & 25.9 & 17.2 & \textbf{12.3} \\
        \cmidrule(){3-9}
        & & \multirow{2}{*}{\texttt{splitwise\_64}} & H20 & 91.8 & 74.3 & 62.7 & 20.4 & \textbf{3.7} \\
        & & & H800 & 69.8 & 51.5 & 29.1 & 26.1 & \textbf{10.7} \\
            
        \midrule 
        \multirow{4}{*}{vLLM} & \multirow{4}{*}{Llama3.1-70B (TP=4,PP=2)} & 
            \multirow{2}{*}{\texttt{arxiv\_16}} & H20 & 69.1 & 45.2 & 54.6 & 0.5 & \textbf{3.0} \\
            & & & H800 & 25.7 & 60.8 & 9.0 & 16.9 & \textbf{0.7} \\

        \cmidrule(){3-9}
        & & \multirow{2}{*}{\texttt{splitwise\_64}} & H20 & 76.7 & 64.7 & 72.6 & 19.1 & \textbf{2.3} \\
            & & & H800 & 49.5 & 67.1 & 38.6 & 23.7 & \textbf{9.9} \\
            
        \bottomrule
    \end{tabular*}
    \vspace{-4mm}
\end{table*}

For single-GPU (TP=1) evaluations, we tested Qwen2.5-14B across all 11 GPUs (Figure~\ref{fig:e2e_accuracy}). \textsc{PipeWeave} achieves an average MAPE of 11.3\%, notably outperforming the best baseline, Neusight at 34.5\%. Furthermore, \textsc{PipeWeave} maintains high accuracy on unseen GPUs, with a 12.5\% MAPE—a significant \textbf{2.8$\times$ }improvement over Neusight's 34.4\%.

This robustness extends to distributed inference. As shown in Table~\ref{tab:e2e_distributed_results}, \textsc{PipeWeave} delivers consistent accuracy across diverse end-to-end inference scenarios. It spans two inference frameworks (SGLang and vLLM), multiple models (Qwen3-32B and Llama3.1-70B), and various parallelism strategies (TP=2, TP=4, TP=8, and TP=4\&PP=2). \textsc{PipeWeave} consistently achieves low MAPE averages: 8.4\% (SGLang, Qwen3-32B, TP=2), 4.3\% (SGLang, Llama3.1-70B, TP=4), 7.7\% (SGLang, Llama3.1-70B, TP=8), and an excellent 4.0\% (vLLM, Llama3.1-70B, TP=4\&PP=2). This performance significantly surpasses the best baseline Neusight. Across all 20 tested configurations, \textsc{PipeWeave} achieves an overall average MAPE of 6.6\% versus Neusight's 34.7\%, showing a \textbf{5.3$\times$} average accuracy improvement.
\rev{Interestingly, our analysis shows that in some E2E inference scenarios, baselines such as Neusight can exhibit very low E2E errors (e.g., 0.5\%) despite having poor kernel-level prediction accuracy. We identify two primary causes for this phenomenon. First, E2E latency aggregates the execution time of many kernels, which leads to systematic error cancellation: overestimations for some kernels offset underestimations for others, thereby reducing the overall E2E error. Second, E2E inference typically involves a much narrower and more constrained set of workload dimensions than those covered in comprehensive kernel-level evaluations (Section~\ref{sec:dataset}); consequently, these workloads often lie near the baseline's prediction ``sweet spots.''}

In summary, \textsc{PipeWeave} delivers \textbf{high fidelity}, \textbf{fast prediction}, and \textbf{broad generalizability} for GPU performance modeling.

\section{Beyond Simulation}
\label{sec:optimization_guidance}

In prior sections, we verified the robustness of \textsc{PipeWeave}. Trained with a MAPE loss, our framework demonstrates strong accuracy in forecasting the performance of various \textit{well-optimized} kernels on diverse hardware platforms.
In this section, we transition from general prediction to a more challenging task: \textbf{optimization guidance}. Our goal is to improve the performance of the Fused MoE Triton kernel---the default MoE backend in SGLang~\cite{sglang}---across hardware platforms. 

The primary challenge lies in the opacity of performance potential. For any given input shape and hardware platform, the attainable performance ceiling is unknown. Consequently, we cannot determine \textit{a priori} whether a current execution is near-optimal or sub-optimal. For instance, achieving 50\% of the roofline\cite{roofline} on an A40 might be poor if the true ceiling is 70\%, while 20\% on an A100 could be near-optimal if the ceiling is only 21\%. \rev{Lacking this ground truth, it is impossible to systematically quantify the performance gap or identify where system-level optimization efforts should be directed.}

Therefore, we look beyond simulation. Rather than predicting average performance, our aim is to provide practical optimization guidance. We aim to address the following questions:
\begin{itemize}
    \item[(1)] Can modeling help establish the kernel's true ``Potential Performance Ceiling'', distinct from the noise of sub-optimal configurations?
    \item[(2)] \rev{Can this estimated ``ceiling'' serve as a reference to identify systematic underutilization and guide optimization efforts?}
\end{itemize}

\subsection{Defining the Potential Ceiling via Quantile Loss}
\label{sec:quantile_methodology}

To address this issue, we adopt the principles of Quantile Regression~\cite{quantile}. We train an MLP model using the same feature set and target (execution efficiency) described in Section~\ref{sec:impl_training}, but employ Quantile Loss as the training objective. We specifically configure the model to predict the 80th percentile (P80). This approach provides a statistically robust estimate of the performance ceiling, which is less sensitive to extreme outliers or measurement noise compared to higher quantiles such as P90.

By targeting P80, the model is effectively trained to fit the top 20\% of performance data points, capturing the characteristics of high-performing configurations while systematically filtering out the lower 80\% of sub-optimal results. Consequently, the resulting prediction, $\hat{y}_{p80}$, does not represent a typical average. Instead, it serves as a statistically-defined \textbf{Potential Performance Ceiling}, representing a high yet realistically achievable target for the kernel's implementation.

\subsection{Diagnosing the Performance Gap}
\label{sec:diagnosing_gap}

We first validate the P80 model as a diagnostic tool. The trained model, which predicts the P80 ceiling $\hat{y}_{p80}$, is applied across the entire Fused MoE dataset (Section ~\ref{sec:dataset}). We then measure the \textit{Performance Gap} by computing the difference between the predicted ceiling and the actual performance:
$$
\text{perf\_gap} = \hat{y}_{p80} - y_{\text{actual}}
$$
Here, $y_{\text{actual}}$ represents execution efficiency (Section~\ref{sec:impl_training}). 

\begin{figure}[tb]
    \centering
    \includegraphics[width=1.0\columnwidth]{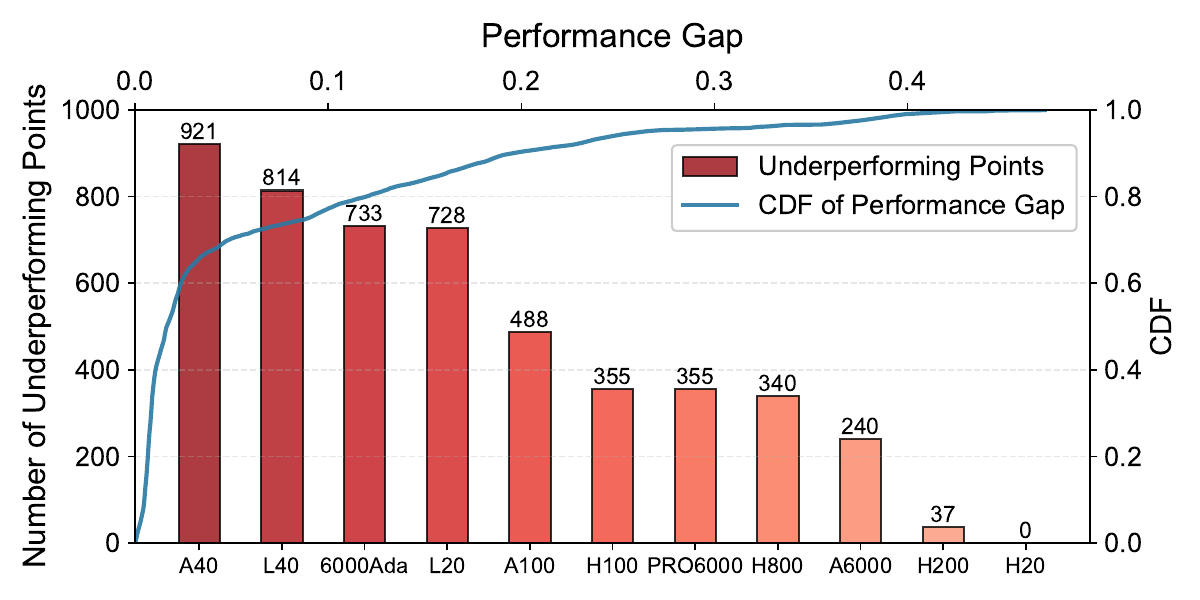}
    \vspace{-8mm}
    \caption{Performance Gap analysis. The CDF of the gap distribution (line) and the count of ``Underperforming Points" (Gap $>$ 0.1) by hardware (bars).}
    \label{fig:gap_analysis_combined}
    \vspace{-4mm}
\end{figure}

Figure~\ref{fig:gap_analysis_combined} presents a consolidated analysis of these gaps.
\rev{Each vertical bar represents a hardware platform, with the bar height indicating the total number of identified underperforming points on that platform. The line plot shows the cumulative distribution function (CDF) of the performance gaps aggregated across all evaluated hardware platforms.}
The figure reveals two key findings. First, the CDF line confirms a ``long tail'' pattern. We observe that while the vast majority of configurations perform near their potential, approximately $80\%$ of all points have a Performance Gap below 0.1. Based on this observation, we identify an \textbf{``Underperforming Point''} as any configuration where the $\text{Performance Gap} > 0.1$. Second, the bar chart pinpoints where these Underperforming Points occur, revealing that significant inefficiencies are hardware-specific. For instance, the A40 GPU exhibits the largest discrepancies, accounting for the vast majority of inefficiencies with \textbf{921} distinct Underperforming Points (representing \textbf{30.4\%} of all A40 samples). This clearly indicates that the kernel's current configuration logic is ill-suited for this specific hardware architecture. In stark contrast, the H20 achieves near-optimal results, exhibiting \textbf{zero} such points.

\subsection{Closing Performance Gap by Tuning Parameters}
\label{sec:closing_gap}

In Section~\ref{sec:diagnosing_gap}, we apply our P80 model to successfully identify ``Underperforming Points''. We now verify that these gaps are actionable \rev{and indicative of systemic optimization potential}. Approximately 70 unique ``Underperforming Point'' configurations are selected for each GPU: A40, L20, A100, and H800. \rev{For these targeted cases, optimization is conducted via brute-force autotuning over three parameters: \texttt{BLOCK\_SIZE}, \texttt{num\_stages}, and \texttt{num\_warps}.}

\rev{To explicitly validate our statistical diagnostic methodology against actual optimization outcomes, Table~\ref{tab:tuning_correlation} shows the relationship between the systemic density of underperforming points and the achieved tuning benefits. A clear positive correlation is observed (Pearson correlation coefficient of \textbf{0.86}): hardware platforms with a higher count of underperforming points obtain larger geometric mean speedups after tuning. This result confirms that our statistical diagnosis effectively reflects real optimization potential and can guide tuning efforts toward configurations with the largest expected gains.}

\begin{table}[htbp]
  \vspace{-1mm} 
  \centering
  \caption{\rev{Speedup vs. underperforming points across GPUs.}}
  \label{tab:tuning_correlation}
  \vspace{-1mm} 
  \begin{tabular*}{\columnwidth}{@{\extracolsep{\fill}}lcc@{}}
    \toprule
    \textbf{GPU} & \textbf{Underperforming Points} & \textbf{Geo-mean Speedup} \\
    \midrule
    A40  & 921 & 1.61$\times$ \\
    L20  & 728 & 1.12$\times$ \\
    A100 & 488 & 1.06$\times$ \\
    H800 & 340 & 1.03$\times$ \\
    \bottomrule
  \end{tabular*}
  \vspace{-1mm} 
\end{table}

\begin{figure}[tb]
    \centering
    \includegraphics[width=1.0\columnwidth]{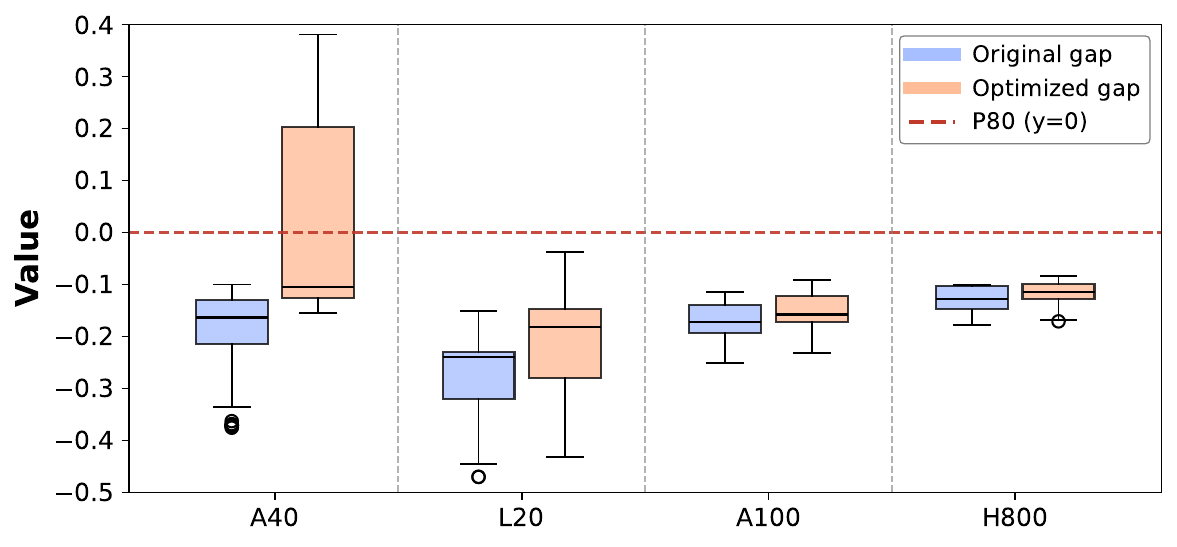}
    \vspace{-7mm}
    \caption{Performance gap distribution before and after model-guided optimization across four GPU platforms.}
    \label{fig:pipetuning}
    \vspace{-5mm}
\end{figure}

Furthermore, Figure~\ref{fig:pipetuning} demonstrates the practical impact of these diagnosed underperforming points. After applying brute-force autotuning, the average performance gap is noticeably reduced, particularly on hardware that initially exhibits larger inefficiencies. For example, the average gap on A40 decreases from \textbf{0.187} to \textbf{0.083}, and on L20 from \textbf{0.274} to \textbf{0.215}. In contrast, the improvements on A100 and H800 are more limited, as their baseline configurations are already closer to the estimated performance ceiling.
Despite these improvements, a residual gap often remains. This suggests that certain inefficiencies cannot be fully eliminated through parameter tuning alone, but may instead stem from deeper factors such as the kernel's structural design or inherent limitations of the Triton programming model~\cite{davis_gpu_programming_models_2025, ringlein_performance_portability_2025}.

\section{Related Work}
\subsection{GPU Performance Modeling}

Research on GPU performance modeling is broadly divided into three categories: cycle-accurate simulators~\cite{gpgpusim,accelsim,mgpusim}, analytical models~\cite{AMALI,GCOM,GPUMech,llmcompass}, and data-driven approaches~\cite{habitat,neusight}. Despite their usefulness, these approaches present inherent trade-offs. High-fidelity cycle-accurate simulators are computationally expensive and difficult to generalize across new hardware. Faster alternatives---analytical and data-driven models---often face limited accuracy, hardware-specific constraints, and coarse-grained assumptions that miss complex behaviors such as fused-kernel coupling, restricting their generalization. \textsc{PipeWeave} is designed to address these limitations by combining principled analytical modeling with the speed and flexibility of data-driven techniques, enabling high fidelity and broad generalization.

\rev{Table~\ref{tab:comparison_model} summarizes representative GPU performance models. Unlike prior methods relying on empirical black-box learning or coarse-grained analytical abstractions (e.g., tile-level throughput and static wave scheduling), \textsc{PipeWeave} advances the grey-box paradigm via a microarchitecture-aware, pipeline-level formulation. By explicitly capturing heterogeneous pipeline demands and dynamic scheduling, it achieves high accuracy and cross-architecture portability.}

\subsection{Network Simulation}
As computation scales across multi-node clusters, precise modeling of data center interconnects grows increasingly vital. General-purpose network simulators~\cite{ns3,omnetpp} provide granular packet-level control to evaluate congestion, routing behavior, and protocol interactions. Newer AI-focused network simulation frameworks~\cite{simai, astrasim} target communication patterns such as All-Reduce and All-Gather, as well as the performance characteristics of large-scale, communication-heavy distributed workloads.

\subsection{System-Level Simulation}
Beyond individual components, extensive research focuses on end-to-end DNN system performance simulation. These tools model the complex interactions among computation, communication, and scheduling strategies for large models. In distributed training, simulators~\cite{astrasim, vtrain, simai} evaluate various parallelism strategies, such as data, pipeline, tensor. For inference, particularly in LLM serving, tools~\cite{llmservingsim, vidur, frontier} simulate dynamic batching and scheduling policies. Our \textsc{PipeWeave} framework not only incorporates a system-level simulator for inference, but also offers a high-fidelity, pluggable GPU computation model required by prior system-level tools.

\begin{table}[t]
\centering
\caption{\rev{Comparison of Microarchitectural Modeling Capabilities.}}
\label{tab:comparison_model}
\renewcommand{\arraystretch}{1.2}
\setlength{\tabcolsep}{4pt}
\resizebox{\columnwidth}{!}{
\begin{tabular}{@{}llll@{}}
\toprule
\textbf{Dimension} & \textbf{Habitat} & \textbf{Neusight} & \textbf{\textsc{PipeWeave} (Ours)} \\ \midrule
\textit{Modeling Strategy} & Black-box & Macro Grey-box & \textbf{Micro-arch Grey-box} \\
\textit{Granularity} & Kernel-level & Tile-level & \textbf{Pipeline-level} \\
\textit{Hardware Fidelity} & GPU & SM & \textbf{Pipeline} \\
\textit{Scheduling Semantics}& N/A & Static wave assumption & \textbf{Dynamic SM scheduling} \\
\textit{Kernel Type} & Elemental kernels & Elemental kernels & \textbf{Fused \& Elemental kernels} \\
\textit{Cross-Arch Generality}& Low & Medium & \textbf{High} \\ 
\textit{Prediction Accuracy} & Low & Medium & \textbf{High} \\
\bottomrule
\end{tabular}
}
\end{table}

\section{Conclusion}
We present \textsc{PipeWeave}, a unified framework that synergizes knowledge-guided analytical modeling with data-driven learning to achieve high-fidelity GPU performance prediction. By decomposing kernels into fundamental pipeline demands and capturing complex runtime interactions via an MLP, \textsc{PipeWeave} demonstrates state-of-the-art accuracy and generalization across diverse kernels, workloads, and hardware generations. Beyond prediction, we validated its practical utility in diagnosing hardware-specific inefficiencies and guiding targeted optimizations.

Future work will focus on two main areas. First, we will extend \textsc{PipeWeave} to complex distributed settings, incorporating support for multi-node clusters and advanced parallelism strategies such as Expert Parallelism (EP). Second, we plan to broaden our model-guided optimization method by developing automated tools that detect performance bottlenecks and enhance configuration logic for more production kernels.

\section{Acknowledgments}
We thank the anonymous reviewers for their constructive feedback and valuable suggestions that improved this work. We also thank our colleagues for helpful discussions. This work used LLMs for text refinement and code generation. This work was supported by Alibaba Tech infra and Reliability Engineering (TRE) in Alibaba Group through Alibaba Innovative Research Program and Alibaba Research Intern Program.

\bibliographystyle{IEEEtranS}
\bibliography{ref}





\end{document}